# Similarity Analysis in Automatic Performance Debugging of SPMD Parallel Programs

Xu Liu[#], Jianfeng Zhan[#], Bibo Tu[#], Ming Zou[#], Dan Meng[#]
[#] *Institute of Computing Technology*
*Chinese Academy of Sciences, Beijing 100190, China*
liuxu@ncic.ac.cn

## ABSTRACT

*Different from sequential programs, parallel programs possess their own characteristics which are difficult to analyze in the multi-process or multi-thread environment.*

*This paper presents an innovative method to automatically analyze the SPMD programs. Firstly, with the help of clustering method focusing on similarity analysis, an algorithm is designed to locate performance problems in parallel programs automatically. Secondly a Rough Set method is used to uncover the performance problem and provide the insight into the micro-level causes.*

*Lastly, we have analyzed a production parallel application to verify the effectiveness of our method and system.*

## 1. INTRODUCTION

Compared to sequential programs, parallel programs are more complex. According to the cause of problems, we categorize the performance problems into two classifications. One is the internal performance problem, which occurs in the local process or thread. For example, poor data locality, poor efficiency in I/O operation and inefficient computing algorithm are all internal problems, occurring in almost all processes and threads. The other is external performance problem such as load imbalance and resource contention, which is caused by the negative competition among different processes and threads. In this paper, we focus on automatically locating and analyzing the external problem existing in SPMD parallel programs without any involvement of users.

For SPMD programs such as MPI and OpenMP, if the program owns high parallelism degree, balanced workload dispatching and resources utilization, the performance behaviors of all processes or threads should resemble each other [1]. Thus based on this observation, we propose a similarity analysis solution to automatic debugging of performance of parallel program.

SCALEA [2] measures the performance similarity of each code region only between any two different processes or threads. Besides, no further analysis is achieved to determine the causes and effects of dissimilarity, both of which are more attractive to users than the dissimilarity itself.

In this paper, we have extended similarity analysis proposed in [2] to locate the performance bottleneck in parallel environment by judging the effects of dissimilarity, and our method is not confined in the one-one similarity analysis. Additionally, in order to extract the micro-level causes of dissimilarity, we introduce the method of Rough Set (RS method) [3], which is used to pinpoint the micro-level cause of dissimilarity and assist programmer to optimize the performance of parallel program.

Contributions in our work can be concluded as two parts: firstly, with the help of clustering method focusing on similarity analysis, an algorithm is designed to locate performance problems in parallel programs automatically. Secondly, a Rough Set method is used to uncover the performance problem and provide the insight into the micro-level causes.

The structure of our paper includes 6 sections. Section 2 summarizes the automatic procedure of program instrumentation. Performance similarity analysis will be presented in section 3. In section 4, RS method is proposed to find the causes of dissimilarity. In section 5, we have chosen a production parallel program to evaluate the effectiveness of the method. Finally we give a conclusion in section 6.

## 2. PROGRAM INSTRUMENTATION

The whole program is instrumented into several code regions which are the minimum units in our next phases of analysis. We provider two means, one is automatic object-code level instrumentation, and the other is interactive source-code level instrumentation.

Automatic instrumentation is based on the compiler we developed. The compiler instruments the code in the phase of parsing. When AST (abstract syntax tree) is built, the compiler could discern the structure of program, such as loops, functions and subroutines. And then code could be inserted into the node of AST. We call this means as object code level instrumentation. Its advantage is that no re-compiling is needed after the instrumentation. However, it is not convenient for users to instrument the code, since it is not users but compiler managing the instrumentation procedure, however controlling the compiler is complex for users, especially for non-expert ones. Thus we design a source code instrumentation method as supplementation.

Source code instrumentation is based on OMPi [4], which is a source-to-source compiler. OMPi can show the program's structure and help users instrument the code. Additionally, we design two interfaces, Par_begin() and Par_end(), for manual instrumentation.

## 3. PERFORMANCE SIMILARITY

Performance similarity is analyzed among all participating computing processes or threads to discover the discrepancy. For SPMD program, each process or thread executes the same code to compute. Thus the high similarity degree in performance of all the executing units means the balance of workload dispatching and resources utilizing and vice versa [1].

We choose CPU time as the main measurement for the performance similarity analysis and other metrics such as cache miss rate, disk I/O quantity and network I/O quantity are accessories to determine the micro-level causes, since CPU time is the most important metric concerned by programmers in determining the performance of a code region. Each process or thread's performance is represented by a vector $\vec{V}_i$ ($i$ is the rank of process or thread), and CPU time of every code region $t, \vec{T}_{it}$ takes up the dimension $t$ in the vector, described as $\vec{V}_i = <T_{i1}, T_{i2} \cdots, T_{in}>$.

Euclidean distance $Dist_{ij}$ is used to judge the similarity between two vectors $\vec{V}_i$ and $\vec{V}_j$, which is calculated in equation 1.

$$Dist_{ij} = \sqrt{(T_{i1} - T_{j1})^2 + \cdots + (T_{in} - T_{jn})^2} \quad (1)$$

We propose a metrics of S to measure the dissimilarity severity of the program in equation 2. Larger S means more severe in performance dissimilarity among processes or threads.

$$S = \frac{\max(Dist_{ij})}{\min(len_i)} \qquad len_i = \sqrt{T_{i1}^2 + ... + T_{in}^2} \quad (2)$$

Based on $Dist_{ij}$ and S, a kind of clustering method-OPTICS [5] is used to classify all the processes or threads. OPTICS deems the performance vector of each process or thread as a point in a multi-dimension space. And a set of points are clustered into one classification when the point density in the area where these points scattered is larger than a threshold. If all processes have similar performance behaviors and only one classification is finally obtained, thus we confirm that the program has good balance in workload dispatching and resources utilizing. Otherwise we presume that performance problems exist in the program and further analysis is needed to locate the causes of imbalance.

We design a top-down algorithm to locate the code regions taking responsibility for the causes of dissimilarity, of which we call *critical code regions* (CCR). We define *l*-CCR as the *critical code region in the lth nested hierarchy* and CCCR as the *core of critical code region*, the most inner CCR on the nested path. CCCR confines the performance problem in a minimum code scope. Figure 1 shows the tree-like structure of the program and we can see the relationship between *l*-CCR and CCCR.

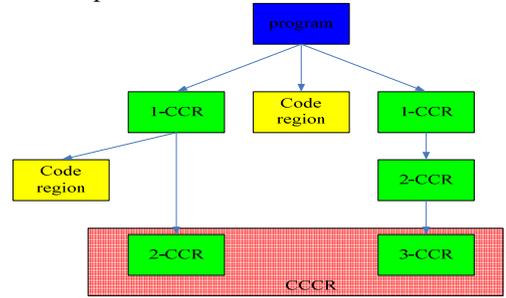

**Figure 1. Hierarchic structure of the program in finding CCCR**

CCCR searching algorithm is explains as follows:

Step 1. If code region $j$'s nested hierarchy is larger than 1, we set $T_{ij}=0$ in vectors for each process $i$.

Step 2. If $T_{ik}>0$, we will watch the effect of changing $T_{ik}$ as 0 on the clustering result - the classification of processes or threads. If we set $T_{ik}=0$ and the classification of processes or threads have changed, then we can confirm the code region $k$ is a CCR, called 1-CCR because it can influence the clustering. If no 1-CCR is found, go to step3, otherwise go to step4.

Step3. We combine n (n=2,3...) code regions into one code region and continue Step1 until finding the 1-CCRs. And the CCCR is the code region which is the intersection of different 1-CCRs. The algorithm ends.

Step4. For each 1-CCR, we continue the top-down analysis according to the nested path. We analyze the *l*th (*l*=2,3,...) nested hierarchy. For each (*l*-1)-CCR, the *j*th code region, in each process $i$, we set $T_{ij}=0$ in the vector and for each code region $k$ nested in it, we set $T_{ik}$ with its CPU time. Re-cluster the vector, if the classification does not change, code region $k$ is a *l*-CCR, because it can represent the characteristic of its outer code region.

Step5. Repeat Step4, until there is no CCR in hierarchy $m+1$. Then *m*-CCR will be added to the core set as CCCR.

## 4. DISSIMILARITY EXTRACTION

Dissimilarity extraction is the further analysis when the processes are clustered into more than one classification. Although we choose CPU executing time as the main criterion in determining the performance, the information it contains is far from comprehensive to provide insight into performance of parallel program. Thus we introduce a Rough Set [3] method to automatically extract the key attributions from accessorial metrics causing the performance dissimilarity.

## 4.1 Introduction to RS

Rough Set is a data mining method that can be used for categorizing, data relating and so on. There are several terms in RS.
- Decision table shown in table 1 is defined to describe the large amount of data. Each entry of the decision is consisted of three parts. One is the ID of entry, one is attribution and another is decision.
- Core is a special set of attributions, which is critical in distinguishing the decisions.

How to find the core and remove the trivial attributions in determining the decision is a main research field in RS. One of the solutions is to create discernibility matrix [6] according to the decision table, the definition of which is as follows: Assume the set of IDs of each entry in the decision table is $\{x_1, x_2, \cdots, x_n\}$. A is the set of attributions and D is the decision. We define a(x) as the value of attribution-a for entry x. From the equation below, we calculate the value of each element $c_{ij}$ in the discernibility matrix.

$$c_{ij} = \begin{cases} \{a \in A : a(x_i) \neq a(x_j)\} & D(x_i) \neq D(x_j) \\ 0 & D(x_i) = D(x_j) \quad i,j = 1 \cdots n \\ -1 & a(x_i) = a(x_j) \quad D(x_i) \neq D(x_j) \end{cases}$$

The discernibility matrix of table is shown in figure 2. Because every discernibility matrix is symmetrical, we only consider its upper triangular part. Using discernibility matrix, extracting the core attributions is much easier and the algorithm is as follows.

**Table 1. An example of decision table**

| ID | a1 | a2 | a3 | a4 | decision |
|----|------|------|------|-------|----------|
| 0  | Sunny | hot | high | False | N |
| 1  | sunny | hot | high | True  | N |
| 2  | overcast | hot | high | False | P |
| 3  | sunny | cool | low | False | P |

$$\begin{pmatrix} 0 & 0 & a1 & a2a3 \\ & 0 & a1a4 & a2a3a4 \\ & & 0 & 0 \\ & & & 0 \end{pmatrix}$$

**Figure 2. The discernibility matrix for the decision table**

Step1. We exclude the elements in the discernibility matrix whose value is 0 or -1 and consider the ones with the value of a set of attributions. If the element owns the value of attribution set which contains only one attribution, we add this attribution into the core set (CS), because it is a core attribution that plays a critical role in making decisions. For example, the CS for figure 2 is {a1}.

Step2. If the value of element in the discernibility matrix, for example {a2a3} in figure 2, does not contain any attributions in the CS, we change the CS as a conjunctive normal form, CS $\wedge$ {a2a3}. In the example, the CS is finally as {a1} $\wedge$ {a2a3} $\wedge$ {a2a3a4}.

Step3. We transform the CS from a conjunctive normal form into a disjunctive normal form. Then we select the conjunctive minor which owns the least number of attributions and occurs in the most times. These attributions are the finally core of the RS. In our example, the core is {a1, a2} or {a1, a3}.

## 4.2 RS in Performance Analysis

In order to utilize the RS method to extract the causes of the dissimilarity of processes' behaviors, our main task is to create decision table and then follow the algorithm depicted in section 4.1, finding the core attributions of RS which are the causes of the dissimilarity.

As shown in Figure 3, we create the decision table according to the three components. The first is entry ID, and we choose the rank of each process to describe. Secondly, we select other metrics except executing time, such as network time and message size, disk I/O time and quantity, L2 cache miss rate and so on, to be the attributions in the decision table. For each attribution, we use OPTICS clustering method to classifying attribution values of different processes and choose the number of classification, of which each process belongs to, as new attribution value to substitute original value of attribution. In order to decrease the data amount and reduce the unnecessary computing, the attributions to be analyzed are confined to the metrics that collected in CCCRs，defined in section 3. Thirdly, the decision value is substituted by the classification number of each process which we obtained in section 3. The rules of creating the decision table is shown in figure 3.

How to extract the core from the attributions by using discernibility matrix method is already depicts in section 4.1. The core set is the attributions that cause the processes behave dissimilarly. Those micro-level causes of dissimilarity of parallel program could help users tune the performance of parallel program.

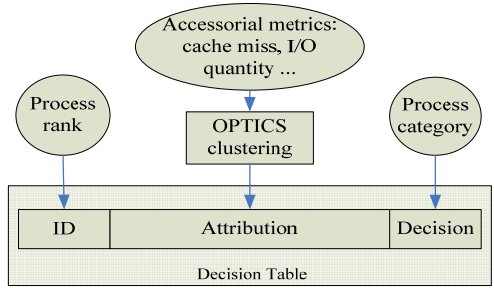

**Figure 3. Creating decision table**

## 5. EVALUATION

Geo-ST is one module of Geo-EAST software developed by Bureau of Geophysical Prospecting (BGP) of China National Petroleum Corporation (CNPC), which is used to calculate seismic tomography using refutations method. Our automatic instrumentation system divided the whole program into 14 regions. The test is done on four SMP machines with 8 CPUs connected with Fast Ethernet 1000Mbps.

The similarity analysis result obtained in section 3 is displayed in figure 4. We can find that all processes are clustered into 5 classifications. Region 11 and 14 are CCRs and region 11 is CCCR. Thus we can draw that the program has external performance problems and it is code region 11 that leads to the problems. From the dissimilarity severity we can see the external problem owns a high severity degree, 0.78. Region 11's workload for each process is dispatched averagely in a static way. In the next phase, we create decision table to analyze the causes of the problems in code region 11.

```
Performance similarity
there are 5 kinds of processes
kind 0: 0
kind 1: 1  2
kind 2: 3
kind 3: 4  6
kind 4: 5  7
dissimilarity severity: 0.783958
CCCR: region 11
CCR tree:
region 14 (1-CCR)  --->  region 11 (2-CCR & CCCR)
```

**Figure 4. The analysis result of similarity measurement**

**Table 2. Decision table for the Geo-ST**

| ID | a1 | a2 | a3 | a4 | a5 | D |
|---|---|---|---|---|---|---|
| 0 | 0 | 0 | 0 | 0 | 0 | 0 |
| 1 | 0 | 0 | 0 | 0 | 1 | 1 |
| 2 | 0 | 0 | 0 | 0 | 1 | 1 |
| 3 | 1 | 0 | 0 | 0 | 2 | 2 |
| 4 | 0 | 1 | 0 | 0 | 3 | 3 |
| 5 | 1 | 1 | 0 | 1 | 4 | 4 |
| 6 | 1 | 2 | 0 | 1 | 3 | 3 |
| 7 | 1 | 2 | 0 | 0 | 4 | 4 |

Table 2 and figure 5 show the decision table and discernibility matrix respectively. In the decision table, the attributions a(i) (i=1,2,3,4,5) respectively represents L1 cache miss rate, L2 cache miss rate, disk I/O quantity, network I/O quantity and executing instruction number, all of which are collected in code region 11. By the analysis of RS, attribution a5, executing instruction of the CCCR is the main reason for the dissimilar behavior of processes. Figure 6 verifies our analysis from which we can discover the obvious difference in executing instruction quantities among different processes running code region 11. Thus we believe that different iterations in the region 11 among different workload leads to the problem.

$$\begin{Bmatrix} 0 & a_5 & a_5 & a_1a_5 & a_2a_5 & a_1a_2a_4a_5 & a_1a_2a_4a_5 & a_1a_2a_5 \\ & 0 & 0 & a_1a_5 & a_2a_5 & a_2a_4a_5 & a_1a_2a_4a_5 & a_1a_2a_5 \\ & & 0 & a_1a_5 & a_2a_5 & a_2a_4a_5 & a_1a_2a_4a_5 & a_1a_2a_5 \\ & & & 0 & a_1a_2a_5 & a_2a_4a_5 & a_2a_4a_5 & a_2a_5 \\ & & & & 0 & a_1a_4a_5 & 0 & a_1a_2a_5 \\ & & & & & 0 & a_2a_5 & 0 \\ & & & & & & 0 & a_4a_5 \\ & & & & & & & 0 \end{Bmatrix}$$

**Figure 5. Discernibility matrix for Geo-ST decision table**

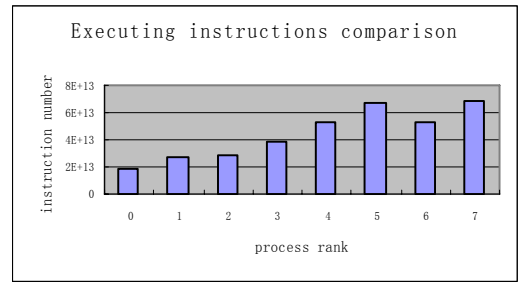

**Figure 6. imbalance in executing instruction of region 11**

We tune the workload in region 11 for each process according to the executing instruction proportion displayed in figure 6, not in an average way as the original. After the optimization, the all computing processes which execute exactly the same quantity of instructions are gathered in the same category, and dissimilarity severity is 0.032800, which means that all processes have the similar performance with balanced workloads and balanced resources utilizing. In order to prove our viewpoint, we have done the sampling to watch the rate of CPU utility of each process. From figure 7 and 8, we can see the optimization has effect on the program's performance. CPU on each node is more efficiently used and the program performance rise to 1.4 times as the original one.

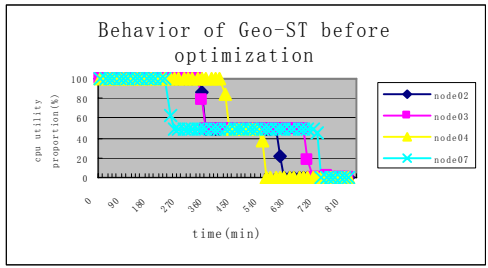

**Figure 7. sampling CPU utility of 4 nodes before optimization**

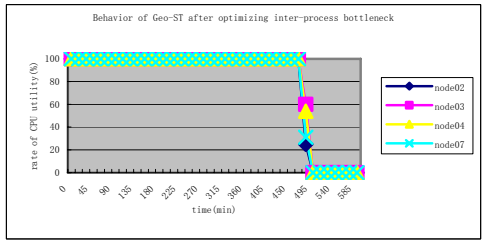

**Figure 8. sampling CPU utility of 4 nodes after optimization**

## 6. CONCLUSION

The paper describes a complete automatic method to analyze the external problems in parallel programs without any users' involvement. There are two main contributions in our work. On the one hand, we extend the similarity analysis to analyze all processes and threads performance, not confined in a one-one comparison. Additionally, an automatic critical code region locating algorithm is proposed to effectively search the location of external problems. On the other hand, we firstly combine the Rough Set method to provide the insight of the micro-level causes of dissimilarity. The experiment can verifies our idea well and achieve a good result.

## 7. ACKNOWLEDGMENTS

This paper is supported by the National Science Foundation for Young Scientists of China (Grant No. 60703020).